**On the giant deformation and ferroelectricity of guanidinium nitrate**

Marek Szafrański[1]* & Andrzej Katrusiak[2]*

[1]Faculty of Physics, Adam Mickiewicz University, Uniwersytetu Poznańskiego 2, 61-614 Poznań, Poland. [2]Faculty of Chemistry, Adam Mickiewicz University, Uniwersytetu Poznańskiego 8, 61-614 Poznań, Poland. Emails: masza@amu.edu.pl; katran@amu.edu.pl

**ARISING FROM**   Karothu et al. *Nature Communications* https://doi.org/10.1038/s41467-022-30541-y

The extraordinary properties of materials accompanying their phase transitions are exciting from the perspectives of scientific research and new applications. Most recently, Karothu et al.[1] described guanidinium nitrate, $[C(NH_2)_3]^+[NO_3]^-$, hereafter GN, as a ferroelectric semiconducting organic crystal with exceptional actuating properties. However, no generally accepted evidence of ferroelectricity and semiconductivity of this hybrid organic-inorganic material was presented. Also, the reproducibility of the large stroke associated with the first-order transition is questionable, because the GN crystals are inherently susceptible to the formation of defects. Moreover, previously published extensive studies on the structure, phase transitions and dielectric properties of GN crystals, i.e. the properties discussed by Karothu et al., were not acknowledged.

In 1992, an unprecedented deformation to about 145% of the original length was reported for the crystals of GN, transforming between their low-temperature phase III and phase II at room temperature around 296 K[2]. In the following years, the crystal structure of GN phase III, built of $D_{3h}$-symmetric ions NH···O hydrogen bonded into honeycomb layers, was determined[3] and the phase transition between phases II and III was extensively characterized by calorimetric, dilatometric, powder X-ray diffraction and dielectric methods[4]. The structures of GN were studied between 4 and 395 K by X-ray and neutron diffraction as well as by neutron inelastic scattering and optical spectroscopy[5-7]. We connected the giant deformation of GN to shear lattice strain and precisely calculated from structural data the elongation of the crystal needle to 144.1% and its contraction across the needle to 68.1%[5]. In the 1990s, these were the largest known deformations between structurally-related polymorphs[8]. Later, a 3-dimensional NH···O bonded high-pressure GN phase IV was revealed and the *p-T* phase diagram including phases I-IV was presented[9]. All in all, eleven GN crystal structures were deposited in the Cambridge Structural Database (CSD), where for phase III we applied an unconventional choice of the unit cell, along the H-bonded layers, to directly relate the lattice dimensions with the deformation of crystals at the phase transition.

In 2022, the unprecedented strain associated with the transition between GN phases III and II was rediscovered by Karothu et al.[1] They cite Szafrański's paper[2] dedicated to the giant deformation of GN crystals, as their reference 48, only in the context of the GN crystallization method, while neglecting the main topic documented by the results of microscopy dilatometry and photographs of the crystal in phases III and II. No other papers[3-7,9] devoted to structural transitions of GN and its dielectric and elastic properties, that is, the main subjects of Karothu's et al. paper, were cited, either.

In the following discussion, we will label the GN phases according to the generally accepted rule, that the highest solid phase before melting is labelled I, and the lower-temperature phases are subsequently numbered[9], *i.e.* differently than in reference 1. Karothu et al. redetermined the structures of GN phases III and II, without mentioning four papers reporting the same crystallographic information before[3,5,7,9]. For phase III, the conventional unit-cell unrelated to the layers[1], such that crystal plane (001) and unit-cell wall (*a*,*b*) lie across the layers of H-bonded ions, prevented them from connecting the deformation of the crystal at the transition to the lattice dimensions. They claim the average elongation of the crystals to 151%, or even more impossible 160% (*cf.* their Supplementary Fig. 10K), which is incompatible with the elongation limit of 144.1% imposed by lattice strain[5]. This latter elongation is consistent with the originally reported[2] dilatometric value of 144.2%. Karothu et al. also claim that the giant magnitude of deformation is retained through at least 20 cycles of the transitions between phases III and II without visible deterioration (their Fig. 2e), whereas even one cycle of the transition leaves structural defects (*cf.* their Supplementary Fig. 10), which accumulate and after very few cycles the macroscopic giant deformation disappears, as it was shown experimentally 30 years ago[2]. The deterioration of the single-crystal quality is clearly seen in the series of subsequent DSC runs presented in Supplementary Fig. 3[1]. Three photographs presented in our Fig. 1a illustrate the typical deterioration of the crystal quality due to the generation of defects at the transition from phase II to phase III, which always reduces the deformation magnitude in subsequent cycles.

Karothu et al. claim that GN is ferroelectric, but they fail to demonstrate the basic ferroelectric properties, like the spontaneous polarization and its switching by an external electric field, or the characteristic temperature dependence of electric permittivity in the paraelectric phase, fulfilling the Curie-Weiss law, or a ferroelectric domain structure below the Curie point[10]. The capacitance-voltage $C(V)$ curves presented in their Fig. 4 are insufficient to prove ferroelectricity. Moreover, a typical $C(V)$ dependence shows a maximum, corresponding to the ferroelectric polarisation switched over in the coercive field ($E_c$)[11,12]. This is due to the $\varepsilon = (1/\varepsilon_0)(\partial P/\partial E)$ relation, showing that the electric permittivity peaks when the polarization is reversed, hence the capacitance should be maximal at $E_c$. Instead, Karothu's et al. observed $C(V)$ plots with the minima, which was not explained nor commented. When hypothetically assuming that these minima appear at the coercive field, the corresponding values of $E_c$ in phase III would be 10–39 V/cm and in phase II 7–29 V/cm. These values could be estimated only, because the dimensions of the sample were not given[1]; on our request, the Authors responded that their samples were 0.1–0.4 mm thick. The $E_c$ values estimated for this thickness are extremely low compared to those of true ferroelectric materials. For example, in guanidinium tetrafluoroborate and guanidinium perchlorate, built from analogous to GN honeycomb N–H···F and N–H···O bonded layers, $E_c$ is 11.4 and 8.7 kV/cm, respectively[13]. The coercive fields of guanidinium aluminum sulfate hexahydrate (1–3 kV/cm), BaTiO$_3$ (0.5–2 kV/cm) and triglycine sulfate (0.43 kV/cm)[10] are slightly lower, but they are still 10 to 100 times higher than the values derived from the measurements by Karothu et al. The atypical shape of $C(V)$ plots and unrealistic values of $E_c$ prompted us to verify the discovery of ferroelectricity in GN. We have tested the GN crystals grown from aqueous[2], water-acetone[1] and water-ethanol solutions, but we detected no differences in their properties. Our Fig. 1b shows that,

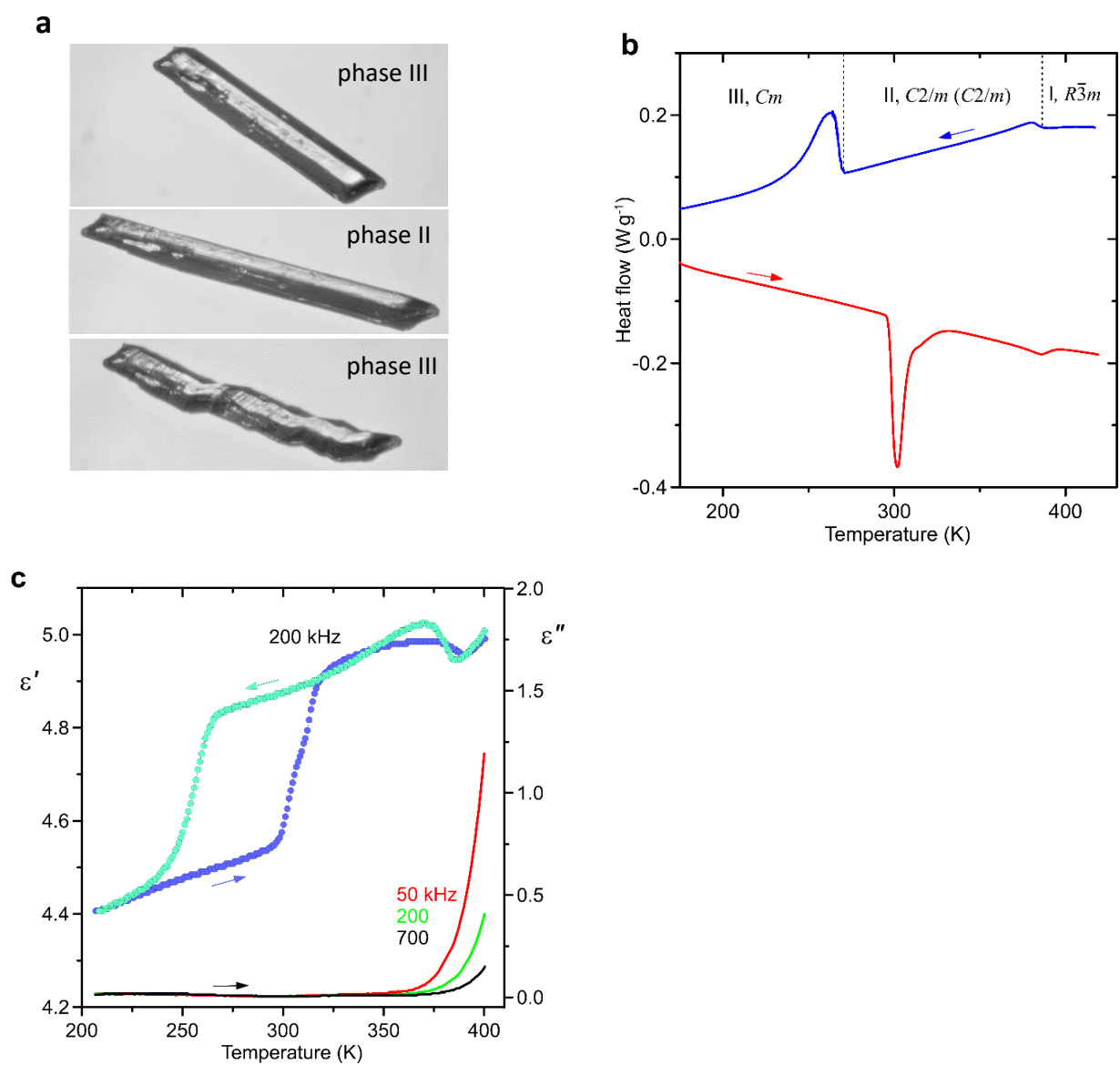

**Fig. 1 Properties of GN at the phase transitions. a** Photographs of a GN crystal as-grown in phase III (top), the same crystal after heating to phase II (middle) and after cooling it again to phase III (bottom). **b** DSC cooling and heating runs measured at the rate 10 K min$^{-1}$ on a Q2000 calorimeter. **c** Real part,

$\varepsilon'$, of the complex electric permittivity $\varepsilon = \varepsilon' - i\varepsilon''$, measured on heating and on cooling the GN sample at a rate 0.5 K min$^{-1}$ (left axis) and the dielectric losses $\varepsilon''$ measured in the heating run (right axis). The polycrystalline GN was pressed into a tablet 0.65 mm thick; silver electrodes were painted on the parallel surfaces, 132 mm$^2$ each.

apart from the first-order transition associated with the huge deformation between phases III and II, GN also undergoes a second-order transition at about 384 K. This is of primary importance, because the high-temperature phase I is centrosymmetric, of space group $R\bar{3}m$[7]. If phase III were ferroelectric, it would follow that one of these two observed phase transitions, either III-to-II or II-to-I, should be of the ferroelectric-paraelectric type. Such a transition type is accompanied by a large anomaly in the electric permittivity and therefore can be easily identified by dielectric measurements, even for polycrystalline samples. Due to the very weak frequency dispersion of the real part of electric permittivity $\varepsilon'$, we have plotted only the data for 200 kHz in Fig. 1c. Both transitions in GN are reflected in the dielectric response of the crystal, but none of them shows any features of the ferroelectric-paraelectric transition. The key test of ferroelectric properties is the spontaneous polarization switching, so we performed it both for a thin polycrystalline film and a single crystal. The linear polarization–electric field dependence and absence of ferroelectric hysteresis loop for the amplitude of the ac electric-field intensity as high as 40 kVcm$^{-1}$, clearly seen in Fig. 2a, testify that GN is non-ferroelectric. Furthermore, our *C*(*V*) tests for the single crystal of GN in phases III and II (Fig. 2b) show that the capacitance value is independent of the electric field, without any butterfly-shaped anomalies. Thus, all of our comprehensive studies detect no traces of ferroelectricity of GN, whatsoever.

In section *Electrical properties*, Karothu et al. describe experiments, where they placed the GN crystals between metal plates made of Ag and Cu; the choice of different electrodes is unfortunate because it usually causes asymmetric results with respect to the zero field[11]. Furthermore, metal plates do not guarantee good contacts with the crystal surfaces, but spurious capacitors can be formed, which can distort the results. Therefore, to perform reliable dielectric measurements, the electrodes should be deposited on the crystal surfaces by sputtering, evaporation, or painting. Besides, a dry atmosphere of measurements is mandatory, since the water adsorbed on the crystal surfaces can dramatically affect the results. None of these requirements was met by Karothu et al. Consequently, they obtained unrealistic high capacitance values for their capacitors containing GN in phases III and II, incompatible with the size of the crystals described in the paper nor with the true values of electric permittivity. The capacitance values of about 6.2 nF in phase III and 55 nF in phase II (their Fig. 4c) juxtaposed with the dimensions of the crystals (their Fig. 1, Supplementary Fig. 10 and Table 5) indicate that the electric permittivity of GN at 200 kHz frequency would have to be in the range of tens of thousands to hundreds of thousands at least, while the true $\varepsilon'$ value is 4.4–5.0 only (our Fig. 1c). We confirmed such relatively low $\varepsilon'$ values, typical for non-ferroelectric dielectrics, also for the single-crystal samples. Noteworthy, the difference in electric permittivity between GN phases III and II amounts to mere 10%, while two orders of magnitude higher change is claimed by Karothu et al. Their huge capacitance reaching thousands nF (*cf.* their Supplementary Table 6), can originate from the high conductivity of their samples exposed to the humid air or from other experimental errors. In Fig. 2c we compare our conductance data measured for single crystal of GN with those reported in their paper. The difference is incredible as it reaches ten orders of magnitude. The 1 kHz conductivity of GN in phase II derived from our data amounts to $2.6 \cdot 10^{-9}$ S/m, which is consistent with the dielectric character of the material, while the value of 17.6 S/m was estimated from the Karothu et al. data (the dimensions of the sample were not specified and therefore we assumed the capacitor thickness 0.1 mm and the electrode area 3 mm$^2$ for this estimation on the basis of the size of crystals presented in their paper). Such a high electric conductivity measured by Karothu et al.[1] is characteristic of semiconductors rather than of insulators. Undoubtedly, this prompted the authors to compare GN crystals with inorganic semiconductors, which is incompatible with the colourless crystal of the absorption edge in the UV region and the energy gap of 5.13 eV (see our Fig. 2d). Moreover, it is well known that the conductance of a capacitor filled with dielectric material increases with the frequency of electric field, which is consistent with our measurements plotted in Fig. 2c, while Karothu et al[1] reported the opposite relationship. Their explanation of the capacitance/conductance decrease by switching off the dipolar and ionic polarization contributions in the frequency range from 1 to 200 kHz is unjustified, because in solid dielectrics the dipolar relaxation occurs in the MHz–GHz frequency range and the resonant frequencies of ionic contributions fall into the infrared region.

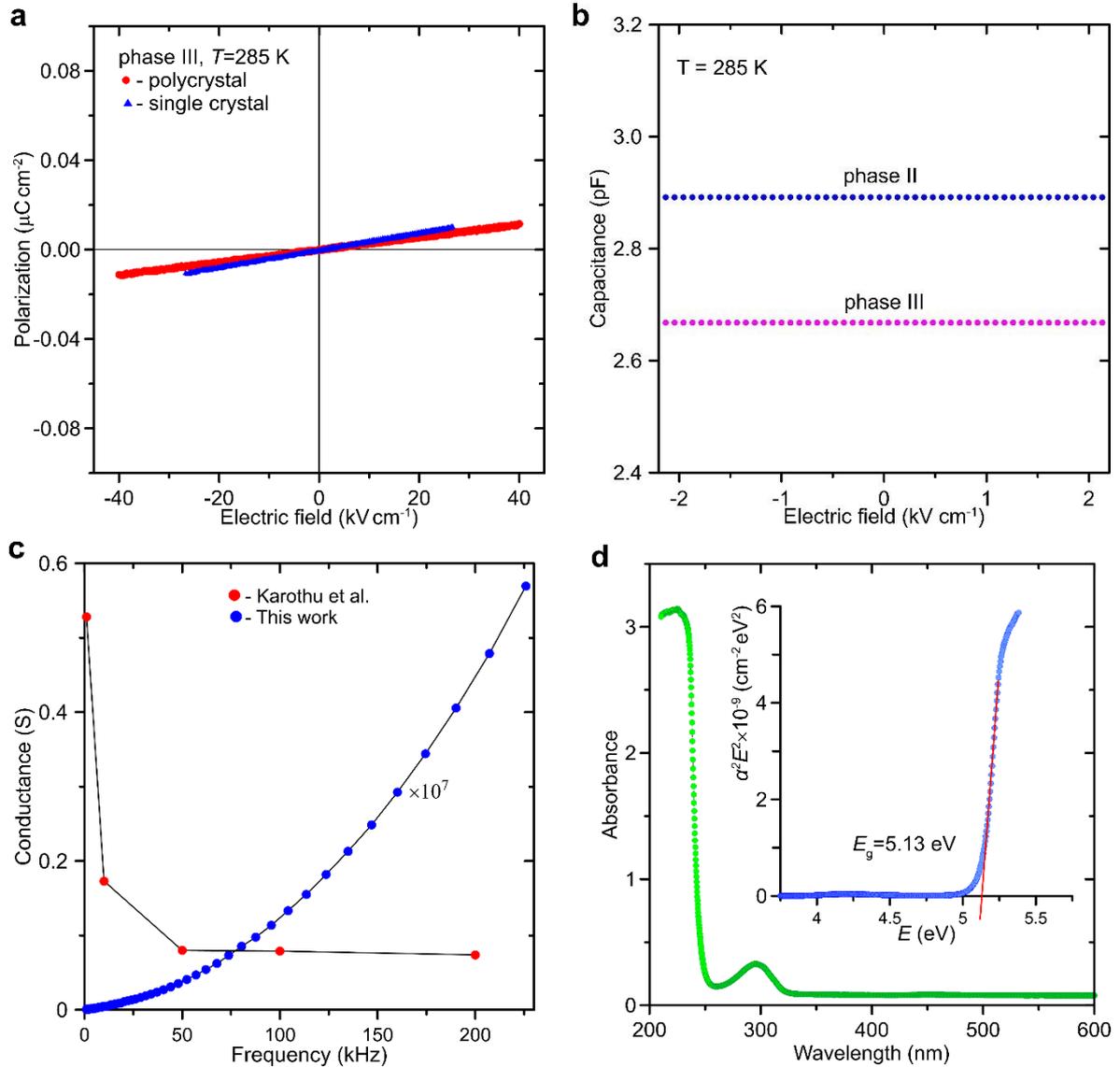

**Fig. 2 Electric conductivity and energy gap of GN. a** Linear polarization-electric field (*P-E*) dependence recorded for GN phase III at 285 K. The measurements were made on the polycrystalline 0.2 mm thick sample with deposited silver electrodes (36 mm$^2$ in area) and on the single-crystal sample (thickness 0.1 mm, surface 3.2 mm$^2$). **b** Capacitance–electric field dependence measured on a single crystal in phases II and III at 285 K. **c** Frequency dependence of the conductance of a capacitor filled with a single crystal of GN, 3.2 mm$^2$ in surface and 0.1 mm thick (blue), compared to the results by Karothu et al. (red): note that our data are multiplied by a factor 10$^7$ to allow their presentation in the same plot. **d** Optical absorption edge of GN. The inset shows the energy gap ($E_g$) determination from Tauc's plot. All preparations and measurements were performed under the dry atmosphere.

In our studies, the low conductivity of GN is confirmed by the very small dielectric losses in both phases III and II around room temperature (our Fig. 1c). The parameters of the capacitors, such as capacitance and conductance presented by Karothu et al., instead of electric permittivity, dielectric loss, and conductivity, are inadequate for characterizing the material properties. Also, incomprehensible are terms 'metal-semiconductor-metal configuration' used for the Ag/GN/Ag system, 'ohmic contact' for a metal/dielectric contact, or 'gate voltage' for the simple plane capacitor. Their paper and its Supplementary Info. contain many other serious errors. For example, their structural mechanism of the phase transition based on the rotations of hydrogen-bonded rings built of three ionic pairs (their Figs. 1g-i, Supplementary Figs. 6 and 11) is unrealistic because of the high energy required for breaking 12 hydrogen bonds and angular momentum connected to such collective rotations. The calculations of energy associated with the rotation of supramolecular rings and a continuous doubling of the unit cell (their Supplementary Fig. 11a-d) cannot be reconciled with the first-order character of the phase transition. Previous $^1$H NMR results[14] show that less energetic reorientations of guanidinium cations around their pseudo $C_3$ axes are activated only in GN phase II above 300 K. Such reorientations require breaking and subsequent formation of six N–H···O hydrogen bonds. Thus, the transition mechanism involving concerted rotations of rings built of six ions tied within the layer with twelve H-bonds, at a much lower temperature in addition, is questionable.

In conclusion, the structural determinations of GN crystals as a function of temperature and pressure, phase transitions, giant deformation, its detailed structural mechanism, molecular dynamics, and dielectric properties of GN were reported before[2-7,9,14]. Undoubtedly, this interesting compound of intriguing properties deserves further studies. However, the citation of previously published results is a generally accepted ethical standard, indispensable for the progress in science. In the 1990s, we did not observe the semiconductivity, ferroelectricity, or fatigue resistance of the giant deformation in GN crystals and our extended measurements presented above do not confirm such properties, either. In the light of our results, the properties of GN crystals are unsuitable for the applications in light-weight capacitors, ferroelectric tunnel junctions, thermistors and multiple-stroke actuators, as proposed by Karothu et al.[1]

**Data availability**

All data obtained and analysed during this study are included in the article.